\documentclass[a4paper,12pt]{article}
\usepackage{tabularx}
\usepackage{amsmath}
\usepackage{amssymb}
\usepackage{graphicx}
\usepackage[margin=0.7in,a4paper]{geometry}
\usepackage[section]{placeins}
\usepackage{mathcomp}
\usepackage{authblk}
\usepackage{url}

\usepackage{gensymb}
\usepackage{float}

\title{\vspace{-1cm} Smallest [5,6]fullerene as building blocks for 2D networks with superior stability and enhanced photocatalytic performance
}
\author[1]{Jiaqi Wu}
\author[2,*]{Bo Peng}
\affil[1]{Peterhouse, University of Cambridge, Trumpington Street, Cambridge CB2 1RD, United Kingdom}
\affil[2]{Theory of Condensed Matter Group, Cavendish Laboratory, University of Cambridge, J.\,J.\,Thomson Avenue, Cambridge CB3 0HE, United Kingdom}
\affil[*]{bp432@cam.ac.uk}
\date{\vspace{-5ex}}

\begin{document}

\maketitle

\begin{abstract}
The assembly of molecules to form covalent networks can create varied lattice structures with distinct physical and chemical properties from conventional atomic lattices. Using the smallest stable [5,6]fullerene units as building blocks, various 2D C$_{24}$ networks can be formed with superior stability and strength compared to the recently synthesised monolayer polymeric C$_{60}$. Monolayer C$_{24}$ harnesses the properties of both carbon crystals and fullerene molecules, such as stable chemical bonds, suitable band gaps and large surface area, facilitating photocatalytic water splitting. The electronic band gaps of C$_{24}$ are comparable to TiO$_2$, providing appropriate band edges with sufficient external potential for overall water splitting over the acidic and neutral pH range. Upon photoexcitation, strong solar absorption enabled by strongly bound bright excitons can generate carriers effectively, while the type-II band alignment between C$_{24}$ and other 2D monolayers can separate electrons and holes in individual layers simultaneously. Additionally, the number of surface active sites of C$_{24}$ monolayers are three times more than that of their C$_{60}$ counterparts in a much wider pH range, providing spontaneous reaction pathways for hydrogen evolution reaction. Our work provides insights into materials design using tunable building blocks of fullerene units with tailored functions for energy generation, conversion and storage. 
\end{abstract}

\section{Introduction}

Carbon atoms provide building blocks for rich structural phases with a variety of physical and chemical properties\,\cite{Esser2017}. The $sp^3$ hybridisation of carbon atoms leads to cubic diamond, one of the hardest materials on Earth; while the covalently bonded $sp^2$ hybridised carbon atoms in honeycomb layers can be held together by van der Waals interactions, resulting in the slipperiness of graphite. Instead of carbon atoms, fullerene molecules can form individual stable units that can be connected through intermolecular bonds, forming superatomic lattices beyond the conventional paradigm of atomic building blocks\,\cite{Zhao2009,Ong2017,OBrien2018,Pinkard2018}. Recently, multiple covalently bonded fullerene networks have been synthesised in 2D, namely, a quasi-tetragonal phase (qTP) and a quasi-hexagonal phase (qHP) in monolayer\,\cite{Hou2022} and few-layer\,\cite{Meirzadeh2023,Wang2023} forms. Such 2D polymeric fullerene is highly stable\,\cite{Peng2022c,Peng2023,Ribeiro2022} with promising electronic and optical properties\,\cite{Hou2022,Tromer2022,Yu2022} for photocatalytic water splitting because of their suitable band gaps and abundant surface active sites on large surface area, which has been predicted theoretically\,\cite{Peng2022c,Jones2023} and soon confirmed experimentally\,\cite{Wang2023}.

Among all the fullerene clusters\,\cite{Kroto1985,Cox1986,Kroto1987}, the (C$_{24}$-$D_{\mathrm{6d}}$)[5,6]fullerene  cage represents the smallest stable conventional fullerene (with 5- and 6-membered rings) that has been predicted theoretically\,\cite{Jensen1998,Manna2016} and characterised experimentally in mass spectra from laser vaporisation products of graphite\,\cite{Hallett1995}. Such synthesis method facilitates fullerene growth through the stacking of C$_{6}$ and C$_{12}$ rings in laser-vapourised hot carbon soot\,\cite{lin_theoretical_2005,kratschmer_solid_1990}, which has also been successfully applied in the synthesis of other fullerene cages such as C$_{36}$\,\cite{piskoti_c36_1998}. The other approach utilises a wide range of organic synthesis techniques to produce a hydrogenated carbon cage\,\cite{ternansky_dodecahedrane_1982,paquette_total_1983}, followed by bromine substitution and gas-phase debromination, which has also been used in the production of fullerene C$_{20}$\,\cite{prinzbach_gas-phase_2000}. Moreoever, C$_{24}$ has been found to be a plausible carrier for the 11.2\,$\mu$m unidentified infrared band in many different galactic and extragalactic environments\,\cite{Bernstein2017}. The highly symmetric molecular structure of C$_{24}$, in combination with the rich variety of carbon-carbon bonds, allow the formation of various monolayer networks of polymeric C$_{24}$ similar to polymeric C$_{60}$. However, it is unclear whether the change of molecular size in such superatomic lattices can tune the physical and chemical properties of monolayer polymeric fullerene in terms of structural stability, electronic structures, optical absorption, and chemical reactivity on the surface. 

In this work, we find that the C$_{24}$ molecules are energetically more favourable to form monolayer networks than C$_{60}$. Additionally, these C$_{24}$ monolayers are more promising as highly stable photocatalysts for overall water splitting. Compared to monolayer polymeric C$_{60}$, monolayer C$_{24}$ networks exhibit superior thermodynamic, dynamic and mechanical stability, indicating the experimental feasibility in synthesising such monolayers. Our hybrid functional calculations show that monolayer phases of C$_{24}$ have comparable band gaps to TiO$_2$, with suitable band gaps to drive overall water splitting over the entire acidic pH range. Most interestingly, while bright excitons near the band edges lead to strong absorption in the solar spectrum for efficient carrier generation, the electron-hole pairs can be effectively separated into individual layers by combining C$_{24}$ with other monolayers in type-II van der Waals heterostructures. Moreover, the large surface area of C$_{24}$ monolayers provides abundant surface active sites to drive the reactions spontaneously in multiple catalytic pathways, with over triple the number of adsorption sites compared to C$_{60}$ monolayers at $\text{pH}>3$.

\section{Methods}

Density functional theory (DFT) calculations are performed using the Vienna \textit{ab initio} Simulation Package ({\sc VASP})\,\cite{Kresse1996,Kresse1996a}. The projector-augmented wave (PAW) basis set\,\cite{Bloechl1994,Kresse1999} is used for C valence states of $2s^22p^2$ under the generalised gradient approximation (GGA) formalism using the Perdew-Burke-Ernzerhof functional revised for solids (PBEsol)\,\cite{Perdew2008}. A plane-wave cutoff of 800\,eV is used with the Brillouin zone sampled by a $\mathbf{k}$-mesh of $5\times5$ and $3\times5$ for qTP and qHP C$_{24}$ monolayers respectively. The lattice constants and atomic coordinates are fully relaxed using the energy and force convergence criteria of $10^{-6}$\,eV and $10^{-2}$\,eV/\AA\ respectively. The interlayer spacing is larger than 23\,\AA, and dipole corrections are applied\,\cite{Makov1995}, in order to eliminate electrostatic interactions across periodic images along $z$. For isolated C$_{24}$ molecules and C atoms, the spacing in all directions is larger than 29\,\AA\ and a $\mathbf{k}$-mesh of $1\times1\times1$ is employed. The atomic coordinates of the C$_{24}$ molecules are fully relaxed with the same energy and force convergence criteria as above. 

The phonon spectrum is calculated under the harmonic approximation using {\sc phonopy}\,\cite{Togo2008,Togo2015} with interatomic force constants computed from density functional perturbation theory (DFPT)\,\cite{DFPT,Gonze1995a}. A supercell size of $2\times2$ and a $\mathbf{k}$-mesh of $3\times3$ are used for both qTP and qHP C$_{24}$. The Helmhotz free energy of the monolayers is evaluated by summing on a $81\times81$ phonon $\mathbf{q}$-mesh and adding electronic contributions. Elastic constants of the monolayers are obtained using the finite difference method\,\cite{LePage2002,Wu2005}.

The electronic structures of the systems are calculated with the hybrid Heyd-Scuseria-Ernzerhof functional revised for solids (HSEsol)\,\cite{Schimka2011} with a screening parameter $\mu=0.2$\,\AA$^{-1}$\,\cite{HSE1,HSE2,HSE3} and PBEsol mixed with unscreened exact Hartree-Fock exchange (PBEsol0)\,\cite{Adamo1999}. It has previously been shown that unscreened hybrid functional PBEsol0 describes best the band gaps of monolayer fullerene networks\,\cite{Peng2022c,Jones2023} compared to the measured ones\,\cite{Hou2022,Wang2023} and the results from the computationally heavy many-body perturbation theory\,\cite{Champagne2024}.

Excitonic effects are calculated from the time-dependent Hartree-Fock (TDHF) method using the Casida equation\,\cite{Sander2017} on top of the PBEsol0 eigenenergies and wavefunctions. Previous work has shown that this method yields quantitative agreements\,\cite{Peng2022c,Jones2023} with the Bethe-Salpeter equation (BSE) on top of many-body perturbation theory\,\cite{Champagne2024}. The Tamm-Dancoff approximation is used as the discrepancy is within 5\,meV\,\cite{Sander2015}. A basis of 16 highest valence bands and 16 lowest conduction bands is used with a $\mathbf{k}$-mesh of $10\times10$ and $6\times10$ for qTP and qHP respectively. The exciton eigenenergies and optical absorption curves are well converged. The dimensionless absorbance in 2D materials is defined as 
\begin{equation}
    A_i(\omega) = \frac{L\omega}{c}\Im{\big[\epsilon_i(\omega)\big]},
\end{equation}
where $\epsilon_i(\omega)$ is the complex dielectric function along polarisation direction $i$ at photon frequency $\omega$, $c$ is the speed of light, and $L$ is the interlayer distance. 

The thermodynamics at different active sites are calculated by placing one hydrogen atom near each symmetry-irreducible carbon atom and then performing full relaxation in a supercell of $3\times3$ (qTP) and $2\times3$ (qHP) to ensure a minimum of $15$\,\r{A} between neighboring adsorbates. Thermal corrections at $T=300$\,K, $p=1$\,atm are added to the electronic free energy by
\begin{equation*}
    \Delta G(T) = \mathrm{ZPE}+\Delta U_{0\rightarrow T}(T)-TS,
\end{equation*}
where ZPE is the zero-point energy, $U$ is the internal energy and $S$ is the entropy, with an extra $pV$ term for gas molecules. Under the standard hydrogen electrode approximation, the free energy of H$_2$ and H$^+$ are equal at equilibrium, while the external potential of photoexcited electrons is taken as the difference between the conduction band minimum and the hydrogen evolution reaction potential\,\cite{Norskov2004,Rossmeisl2007}. 

\section{Results and Discussion}

\subsection{Crystal structures}

\begin{figure}[ht] 
        \centering \includegraphics[width=0.5\columnwidth]{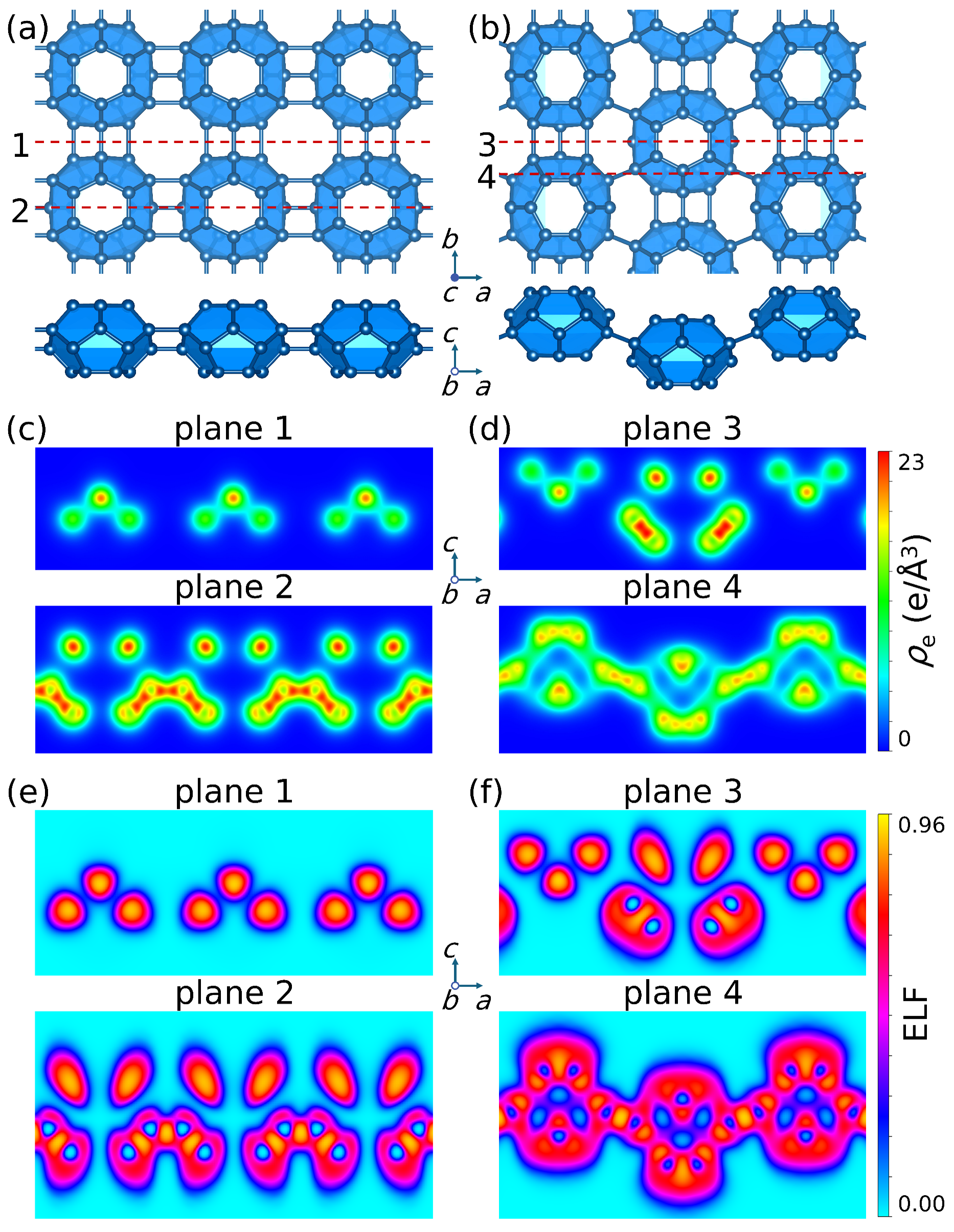} 
        \caption{Top and side views of crystal structures of (a) qTP and (b) qHP C$_{24}$ monolayers. 
        Charge density on selected (010) planes plotted by {\sc VESTA}\,\cite{vesta} showing the intermolecular bonding features of (c) qTP and (d) qHP C$_{24}$ monolayers, as well as their corresponding ELF for (e) qTP and (f) qHP C$_{24}$ monolayers.  }
        \label{fig:elec}
\end{figure}

The crystal structure of the qTP and qHP monolayers are presented in Figure\,\ref{fig:elec}(a) and (b). The overall structure for C$_{24}$ qTP monolayer is analogous to the C$_{60}$ equivalent and can be regarded as a nearly-square lattice with intermolecular bonds between neighbouring C$_{24}$ units. One difference between qTP C$_{24}$ and C$_{60}$ monolayers is that the lattice constants $a$ and $b$ are equal for C$_{24}$ but slightly different for C$_{60}$ (Table\,\ref{tbl:Eb}) because of different molecular symmetry. Moreoever, due to the different molecular geometry, qTP C$_{24}$ monolayers exhibit three non-coplanar bonds between neighbouring C$_{24}$ clusters, instead of the [2+2] cycloaddition bonds for qTP C$_{60}$. In addition, the central bond is slightly shorter than the two side bonds, leading to higher charge density, as shown on plane\,1 in Figure\,\ref{fig:elec}(c). This suggests stronger cohesive interaction relative to their C$_{60}$ counterparts.

\begin{table}[ht]
    \centering
    \begin{tabular}{ccccc}
    \hline
    phase                & $a$ (\AA)                    & $b$ (\AA)                    & $E_\text{c}$ (eV)   & $\Delta E_\text{c}$ (eV)     \\ \hline
    0D & - & -              & $-8.586$   & {0.000} \\
                         & - & -                               & ($-9.256$) &    (0.000)          \\
    2D qTP & 6.103  & 6.103 & $-8.914$   & $-0.328$           \\
                         &   (9.097) & (9.001)      & ($-9.259$) & ($-0.002$)         \\
    2D qHP & 11.500 & 6.180 & $-8.974$   & $-0.388$           \\
                         &   (15.848) & (9.131)   & ($-9.246$) & (+0.010)          \\ \hline
    \end{tabular}
    \caption{Lattice constants, cohesive energies per atom $E_\text{c}$ and additional cohesive energy for monolayer formation from molecules $\Delta E_\text{c}$ of C$_{24}$ phases calculated from PBEsol. The values for C$_{60}$ are shown in parenthesis for comparison. } 
    \label{tbl:Eb} 
\end{table}

The qHP C$_{24}$ monolayer can be interpreted as misaligned 1D chains along the $b$ direction connected by the three non-coplanar bonds, which are further joined through diagonal single bonds between neighbouring chains along $a$. The monolayer exhibits a buckled structure, owing to the asymmetry of the interchain bonding positions. The charge density plots on plane\,3 in Figure\,\ref{fig:elec}(d) show that the three non-coplanar bonds along $b$ are similar to the qTP bonds, while the diagonal bonds are slightly weaker. This leads to mechanical anisotropy as discussed later. We also investigate the electron localization function (ELF)\,\cite{Becke1990,Savin1992,Gatti2005}. ELF = 1 corresponding to perfect localisation and ELF = 0.5 corresponding to the electron-gas like pair probability. The ELF plots in Figure\,\ref{fig:elec}(e) and (f) show large areas of ELF = 0.5 around the C$_{24}$ units. This indicates the diffuse, delocalised $\pi$ features of the C$_{24}$ cage.

The cohesive energy per atom is defined as
\begin{equation}
    E_\text{c} = E_\text{monolayer}/n-E_\text{atom},
\end{equation}
where $E_\text{monolayer}$ is the energy of the monolayer per unit cell, $n$ is the number of carbon atoms in the unit cell, and $E_\text{atom}$ is the energy of an isolated carbon atom. As summarised in Table\,\ref{tbl:Eb}, both monolayers are more stable with respect to isolated C$_{24}$ molecules by 0.328\,eV/atom for qTP and 0.388\,eV/atom for qHP. Monolayer qHP C$_{24}$ is energetically more favourable by 0.060\,eV/atom, possibly due to its close-packed structure. In comparison, the additional cohesive energy for formation of monolayer C$_{60}$ networks is much weaker, with $-0.002$\,eV for qTP and (endothermic) $+0.010$\,eV for qHP. This can be rationalised by the release of stereochemical strain in C$_{24}$ units through forming $sp^3$-like sites at the non co-planar bonds, suggesting energetically more favourable formation of C$_{24}$ monolayers from molecules. 

\subsection{Stability}

\begin{figure}[ht] 
        \centering 
        \includegraphics[width=0.5\columnwidth]{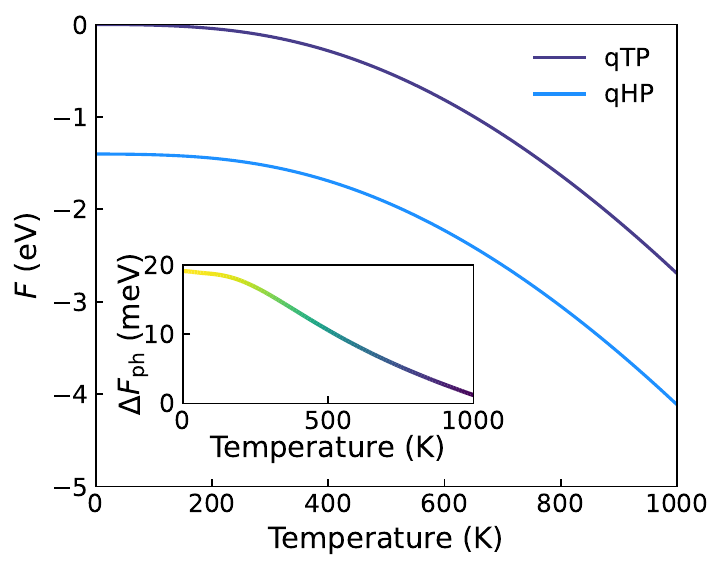} 
        \caption{Free energy curves of qHP and qTP phases per C$_{24}$ unit, with the free energy of qTP at 0\,K set to zero. The phonon free energy difference between qTP and qHP ($\Delta F_{\text{ph}}$) is shown in the inset.
        } 
        \label{fig:thermo}
\end{figure}

\begin{figure*}[ht] 
        \centering 
        \includegraphics[width=\columnwidth]{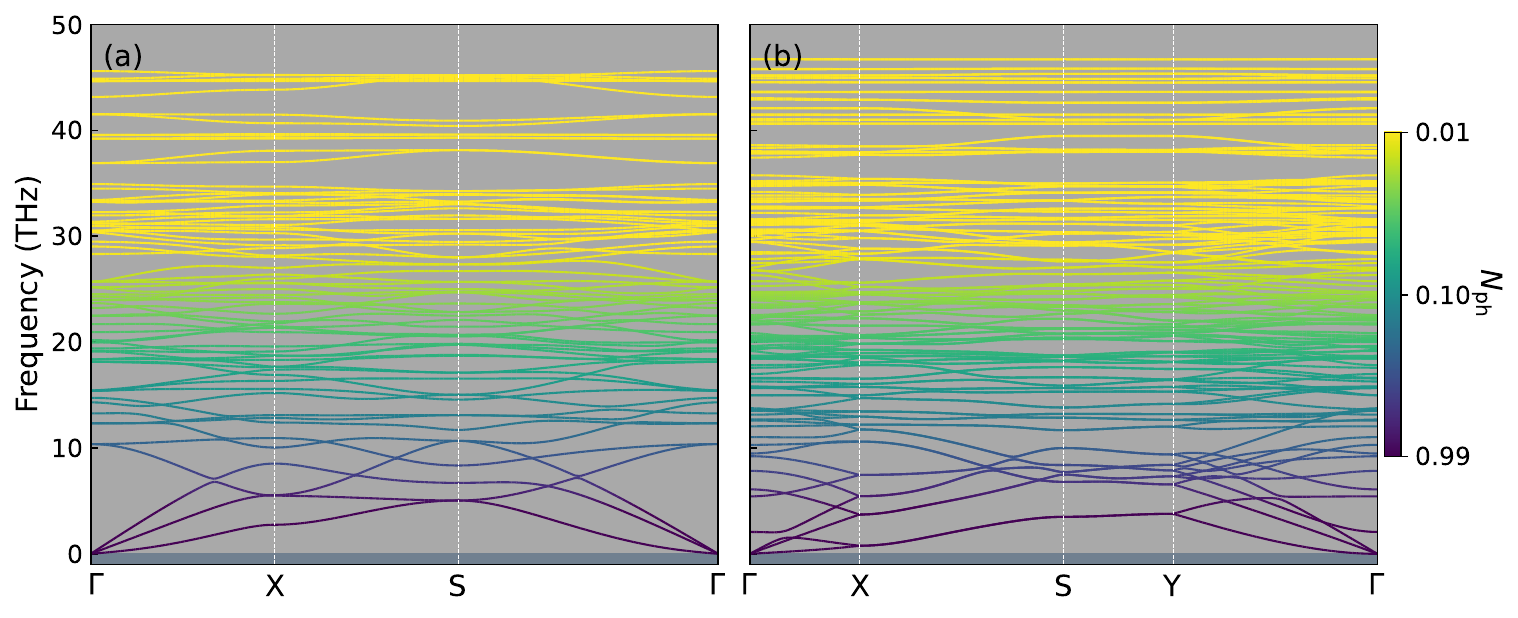} 
        \caption{Phonon spectra of (a) qTP and (b) qHP C$_{24}$ monolayers, with the phonon occupation number $N_{\text{ph}}$ from the Bose-Einstein distribution at 300\,K. } 
        \label{fig:phn}
\end{figure*}

The thermodynamic stability of the monolayers is analysed by considering electronic and vibrational free energies. The variation of total free energy $F$ with temperature is shown in Figure\,\ref{fig:thermo}. The temperature dependence in $F$ is due to the thermal activation of phonon modes shown by the Bose-Einstein phonon occupation number in Figure\,\ref{fig:phn}, while the approximately constant difference of total free energy between qTP and qHP is dominated by the electronic free energy difference of $\Delta F_\text{el} = 1.42$\,eV. The phonon free energy for qHP is slightly higher at low temperatures due to the zero-point energy of more high-frequency phonon modes, and the difference $\Delta F_\text{ph}$ approaches zero around 1000\,K.
The thermodynamically stable phase at all temperatures of interest is qHP, in contrast to the results for C$_{60}$ monolayers\,\cite{Peng2023}. 

To evaluate the dynamic stability of the C$_{24}$ monolayers, the phonon dispersion curves are investigated in Figure\,\ref{fig:phn}. Near $\Gamma$, both dispersion curves exhibit two Debye-like linear acoustic modes corresponding to the in-plane longitudinal and transverse modes. The additional quadratic acoustic mode corresponds to out-of-plane flexural mode, which is a characteristic of monolayer structures. Both phases exhibit no imaginary-frequency phonon mode, indicating that the system is at an energy minimum on the potential energy surface\,\cite{Stoffel2010,Pallikara2022,Huo2024}. The dynamic stability is a sign for strong covalent bonding of the C$_{24}$ monolayers. 

We next confirm the mechanical stability of the two monolayer phases. The 2D elastic constants (in Voigt notation) are normalised from the 3D lattice constants by
\begin{equation}
    C_{ij}^\text{2D} = C_{ij}^\text{3D} \times L,
\end{equation}
where $L$ is the interlayer separation. The elastic constants are summarised in Table\,\ref{tbl:elastic}. The Born-Huang dynamical lattice theory sets the criteria of mechanical stability for qTP (space group $P\Bar{4}m2$) 
\begin{equation}
    C_{11},C_{22},C_{66}>0,\ \ C_{11}+C_{22}+2C_{12}>0,
\end{equation}
and an additional criterion for qHP (space group $Pmmn$)
\begin{equation}
    C_{11}+C_{22}-2C_{12}>0.
\end{equation}
Both qTP and qHP fit these criteria, indicating good mechanical stability. 

\begin{table}[ht]
    \centering
    \begin{tabular}{cccccccc}
    \hline
    phase & $C_{11}$   & $C_{22}$   & $C_{12}$  & $C_{66}=G^\text{2D}$   & $\gamma$ & $Y_{a}^\text{2D}$    & $Y_{b}^\text{2D}$        \\ \hline
    qTP   & 239.9 & 239.9 & -6.2 & 78.2  & 116.8 & 239.7 & 239.7  \\
    qHP   & 221.3 & 268.9 & 20.6 & 103.4 & 132.8 & 219.7 & 267.0  \\ \hline
    \end{tabular}
    \caption{Elastic constants $C_{ij}$, layer moduli $\gamma$, Young's moduli $Y^\text{2D}$ and shear moduli $G^\text{2D}$ (N/m) of qTP and qHP. } 
    \label{tbl:elastic} 
\end{table}

The elastic moduli can be calculated using 
\begin{align}
    \gamma &= \frac{1}{4}(C_{11}+C_{22}+2C_{12}),\\
    Y^\text{2D}_{a} &= \frac{C_{11}C_{22}-C_{12}^2}{C_{22}},\\
    Y^\text{2D}_{b} &= \frac{C_{11}C_{22}-C_{12}^2}{C_{11}},\\
    G^\text{2D} &= C_{66},
\end{align}
where $\gamma$ is the layer modulus (the 2D equivalent of the bulk modulus), and $Y^\text{2D}$ and $G^\text{2D}$ are the 2D Young's and shear moduli respectively. Monolayer qTP C$_{24}$ is elastically isotropic along the $a$ and $b$ directions due to its $S_4$ (four-fold improper rotation) symmetry. In monolayer qHP C$_{24}$, $Y_{a}^\text{2D}$ is smaller than $Y_{b}^\text{2D}$, in agreement with the weaker diagonal interchain single bonds.  In comparison, qHP C$_{24}$ has a larger $\gamma$ and $Y_{b}^\text{2D}$ than qTP due to the more close-packed structure and higher concentration of inter-fullerene bonds. The shear modulus $G^\text{2D}$ of qHP is much larger than qTP due to the difficulty of dislocating the close-packed misaligned chains. Interestingly, the Young's and layer moduli of both C$_{24}$ monolayers are about 1.5 times higher than their C$_{60}$ counterparts\,\cite{Peng2023}. This is rationalised by the presence of triple non co-planar bonds in C$_{24}$ monolayers, compared with the [2+2] cycloaddition bonds of C$_{60}$ monolayers. Another contributing factor is the smaller molecular size and hence higher density of inter-fullerene bonds. 

Given its superior stability and strength comparing to monolayer polymeric C$_{60}$, it is plausible that synthesis and exfoliation of the C$_{24}$ monolayers are experimentally more feasible. We next explore the photocatalytic properties of monolayer C$_{24}$ because of their large surface area for maximum contact with aqueous species for reaction. The crucial requirements for a photocatalyst are (i) a suitable band edge for driving the reaction, (ii) strong optical absorbance to generate photoexcited carriers effectively, and (iii) abundant surface active sites, which will be discussed in the next three subsections. 

\subsection{Electronic structures}

\begin{figure*}[ht] 
        \centering 
        \includegraphics[width=\columnwidth]{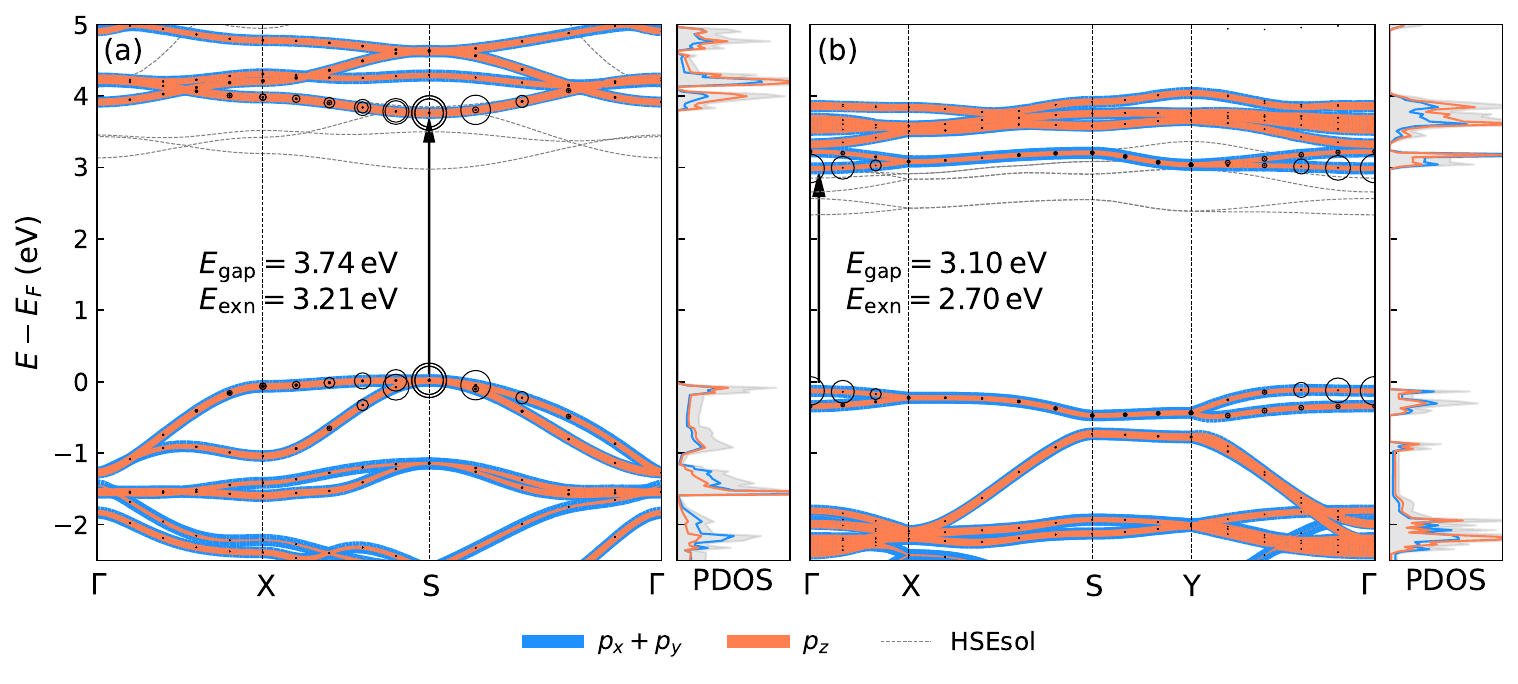} 
        \caption{Orbital-projected band structures with projected density of states (PDOS) for (a) qTP and (b) qHP calculated at PBEsol0 level. The HSEsol band structures are shown in light grey for comparison. The electron-hole pairs that contribute to the first bright exciton are also shown, with larger circles indicating more contributions from the electron-hole pairs at a given $\mathbf{k}$. } 
        \label{fig:band}
\end{figure*}

Figure\,\ref{fig:band} shows the band structure of qTP and qHP C$_{24}$ monolayers calculated from unscreened hybrid functional PBEsol0. PBEsol0 predicts the most accurate band gap for monolayer and few-layer polymeric C$_{60}$\,\cite{Peng2022c,Jones2023} that is in good agreement with the measured band gap\,\cite{Hou2022,Wang2023} and with the many-body perturbation theory\,\cite{Champagne2024}, hence we focus on the PBEsol0 band edges hereafter. Monolayer qTP C$_{24}$ has a direct band gap of 3.74\,eV at the S high-symmetry point. Monolayer qHP C$_{24}$ has an indirect band gap of 3.10\,eV, with the valence band maximum (VBM) around the middle of $\Gamma$-Y and the conduction band minimum (CBM) at $\Gamma$. However, the energy difference between VBM and the highest valence band at $\Gamma$ is less than 14\,meV, leading to a direct-like characteristic. Interestingly, the PBEsol0 band gaps of qTP and qHP C$_{24}$ monolayers are comparable to those of TiO$_2$, the most widely used oxide for photocatalytic applications\,\cite{Fujishima1972,Deak2011,Scanlon2013,Pfeifer2013,Ju2014,Mi2015,Zhang2015n,Deak2016,Chiodo2010,Li2020a,Brlec2022}. 
As a comparison, HSEsol leads to smaller band gaps of 3.03 and 2.42\,eV for qTP and qHP C$_{24}$ respectively, providing the lower bound of the band gaps. The HSEsol valence bands overlap with the PBEsol0 bands when the Fermi energy is set to zero, while the HSEsol conduction bands are a rigid shift of the PBEsol0 bands.

\begin{figure}[ht] 
        \includegraphics[width=\columnwidth]{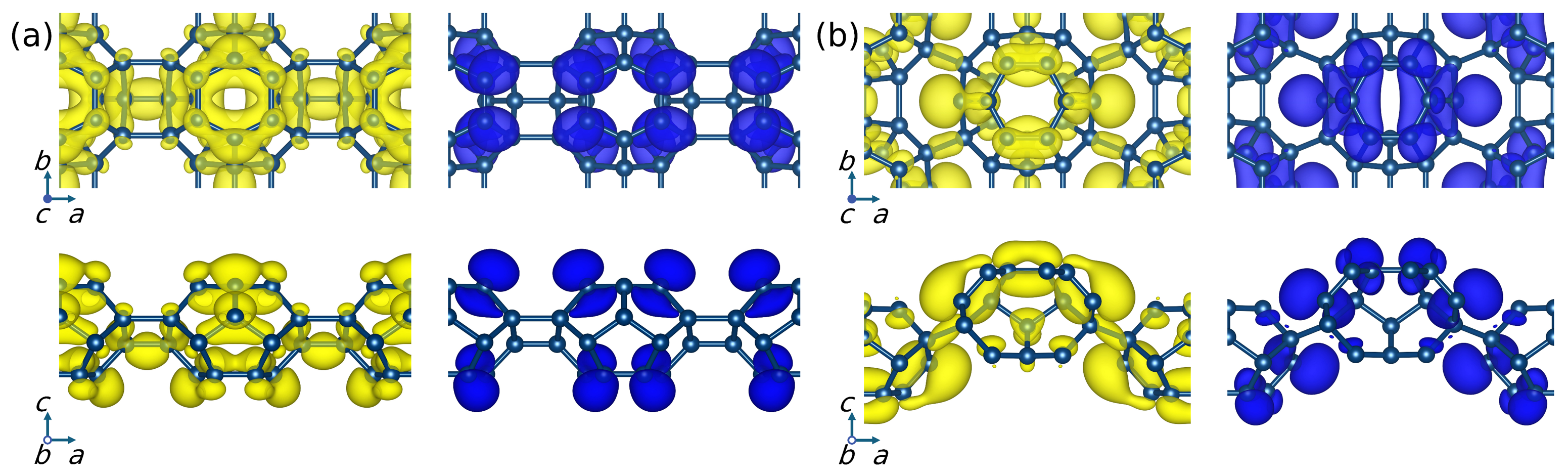} 
        \caption{Top and side views of the VBM (yellow) and CBM (navy) charge densities with the largest contribution to the first bright exciton for (a) qTP and (b) qHP C$_{24}$. The VBM of qTP C$_{24}$ corresponds to a superposition of doubly degenerate states.} 
        \label{fig:charge}
\end{figure}

The band edges are mainly contributed by the $2p$ orbitals of carbon. The $p_z$ orbitals consist of half the DOS at the band edge, resulting in highly delocalised states. Figure\,\ref{fig:charge}(a) and (b) show the band edge states of qTP and qHP respectively. The VBM states of qTP is doubly degenerate at S with significant contributions from the $\pi$ orbitals from the top and bottom hexagonal rings of C$_{24}$ units. The CBM of qTP is a combination of two $\pi^*$ orbitals from the top hexagonal ring and two $\pi^*$ orbitals from the bottom hexagonal ring, in line with the almost full $p_z$ character. For qHP, the partial charge density mainly comes from the $\pi$ and $\pi^*$ orbitals of the hexagonal ring on the outward side of the buckled structure for both VBM and CBM. The delocalised band edge states on the top and bottom hexagonal rings can be promising for stereochemical interactions between the catalyst and aqueous species. 

\begin{figure}[ht] 
        \centering 
        \includegraphics[width=0.5\columnwidth]{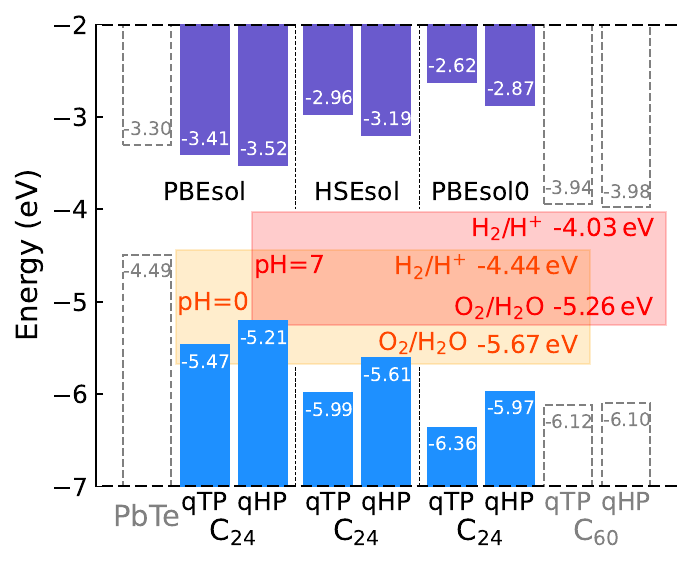}
        \caption{Band edges of monolayer qTP and qHP C$_{24}$ with respect to vacuum level calculated from PBEsol, HSEsol and PBEsol0 compared with the redox (HER/OER) potentials of water at $\text{pH}=0$ and 7. The purple and blue rectangles represent the CBM and VBM respectively. } 
        \label{fig:pht}
\end{figure}

For a water-splitting photacatalyst, the band edges need to straddle the redox potentials of water, i.e., the VBM must be lower than the oxygen evolution reaction (OER) potential of water, $-5.67+\text{pH}\times0.059\,\text{eV}$, and the CBM must be higher than the hydrogen evolution reaction (HER) potential of water, $-4.44+\text{pH}\times0.059\,\text{eV}$. Figure\,\ref{fig:pht} shows the VBM and CBM calculated from PBEsol, HSEsol and PBEsol0, with the vacuum level set to zero under the absolute electrode potential model. The redox potentials of water are shown as horizontal rectangles at $\text{pH}=0$ and 7.

Benefiting from the large band gaps, both qTP and qHP C$_{24}$ fit the photocatalysis criteria quite well even at the PBEsol level in a wide pH range. We hereafter focus on the PBEsol0 band edges because unscreened hybrid functional is more accurate in describing systems with weak screening\,\cite{Savory2016a}. For qTP C$_{24}$, the VBM is 0.69\,eV below the OER potential at $\text{pH}=0$ and the CBM is 1.41\,eV above the HER potential at $\text{pH}=7$. For qHP C$_{24}$, the VBM is 0.30\,eV below the $\text{pH}=0$ OER potential, and the CBM is 1.16\,eV above the $\text{pH}=7$ HER potential. Therefore, both qTP and qHP C$_{24}$ can be promising photocatalysts in a wide pH range from acidic to neutral pH. In contrast, C$_{60}$ monolayers have much smaller band gaps, with the CBM almost touching the HER potential at $\text{pH}=7$\,\cite{Peng2022c}. This suggests enhanced photocatalytic activity of C$_{24}$ monolayers comparing to monolayer C$_{60}$ networks.

\subsection{Optical properties}

\begin{figure*}[ht] 
        \centering 
        \includegraphics[width=\columnwidth]{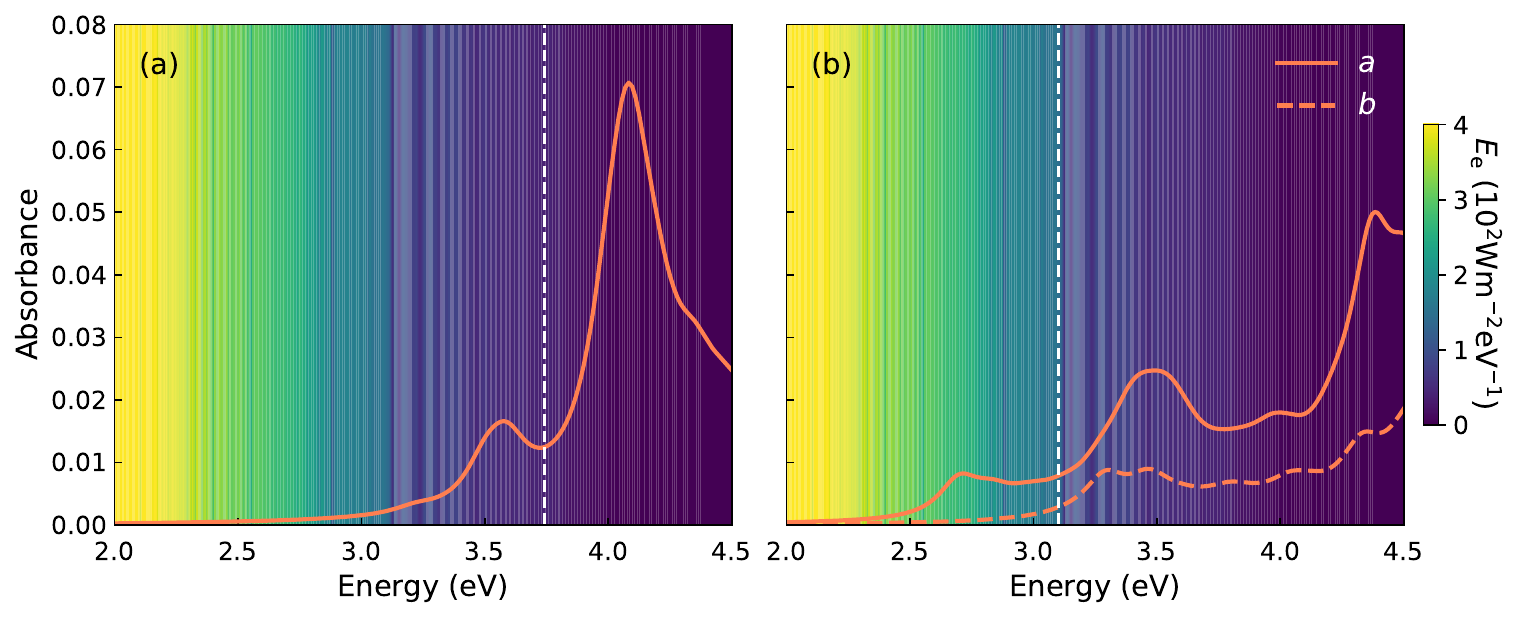} 
        \caption{Excitonic absorbance of (a) qTP and (b) qHP C$_{24}$ monolayers with the band gaps shown as the white lines. For qHP C$_{24}$, the solid and dashed lines correspond to the $a$ and $b$ polarisation directions. The background shows the global total spectral irradiance $E_{\rm{e}}$ from the Sun. } 
        \label{fig:opt}
\end{figure*}

We then consider optical absorbance of C$_{24}$ monolayers to investigate whether enough photoexcited carriers can be generated under solar spectrum. Figure\,\ref{fig:opt} shows the exciton absorbance of C$_{24}$ monolayers computed from PBEsol0\,+\,TDHF. Both phases exhibit a strong exciton absorption below the band gap (white dashed line) due to the direct nature of band gap and suitable molecular symmetry for optical transition selection rules. For qTP C$_{24}$, the first bright exciton has an eigenenergy of 3.21\,eV with an exciton binding energy of 0.53\,eV, which is contributed mainly by the VBM and CBM states at S, as shown by the exciton fatband in Figure\,\ref{fig:band}(a). Even brighter excitons with higher oscillator strength at 3.50 and 3.59\,eV result in strong absorption peaks in the long-wavelength ultraviolet (UV-A) region. 

For qHP, the absorbance is highly anisotropic along $a$ and $b$. There is a strong exciton absorption peak above 2.5\,eV along $a$, corresponding to the first bright exciton of 2.70\,eV with a binding energy of 0.40\,eV consisting of the VBM and CBM states at $\Gamma$. The absorbance along $b$ is much weaker than that along $a$ below the band gap. Above the band gap, strong absorbance peaks are found around 3.5\,eV along both directions, which further enhance UV-A absorption. 

To separate electrons and holes effectively, C$_{24}$ monolayers can be combined with other 2D materials to form van der Waals heterostructures with type-II band alignment. As an example, monolayer PbTe in a $4\times4$ supercell only has a slight lattice mismatch below 0.5\% with qTP C$_{24}$ in a $3\times3$ supercell, and their band edges form the type-II alignment (Figure\,\ref{fig:pht}). On the other hand, the type-I band alignment between qTP and qHP monolayers, in combination with the strongly bound excitons with large oscillator strength and high binding energy, can be utilised in light-emitting devices owing to high emission efficiency\,\cite{Zheng2018}. 

\subsection{Surface active sites}

\begin{figure*}[ht] 
        \centering 
        \includegraphics[width=\columnwidth]{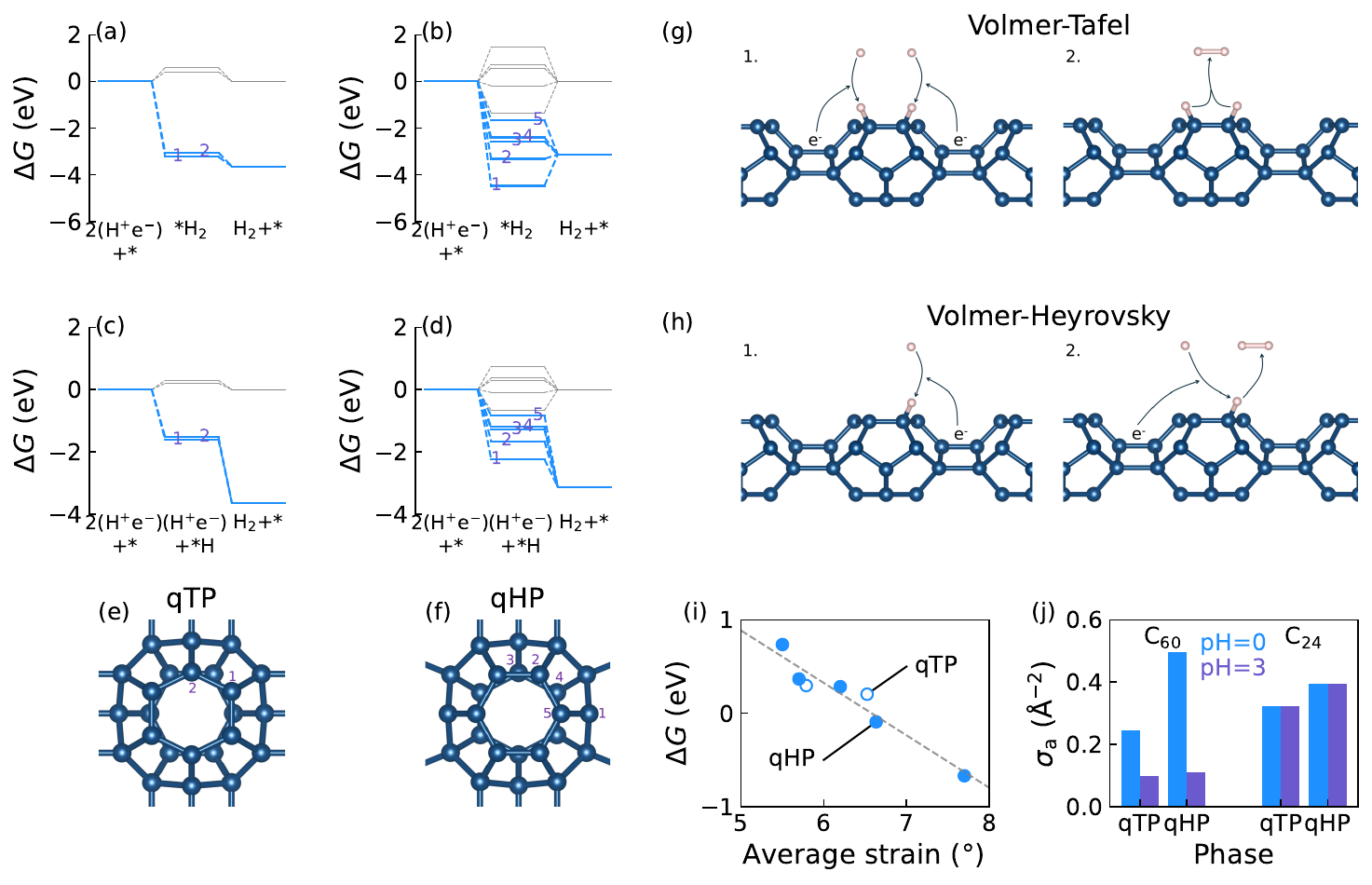} 
        \caption{Free energy profiles of hydrogen evolution at different adsorption sites through the Volmer-Tafel mechanism for (a) qTP and (b) qHP C$_{24}$ and the Volmer-Heyrovsky mechanism for (c) qTP and (d) qHP C$_{24}$ at $\text{pH}=0$, with grey and blue lines representing the absence and presence of photoexcitation respectively. Symmetry-irreducible adsorption sites for (e) qTP and (f) qHP C$_{24}$. (g) Volmer-Tafel and (h) Volmer-Heyrovsky reaction mechanisms. (i) Correlation between adsorption free energy and average bond angle strain. (j) Area density $\sigma_\mathrm{a}$ of active sites for various phases of fullerene monolayers at changing pH.  
        } 
        \label{fig:site}
\end{figure*}

The thermodynamic driving force for water splitting is investigated by calculating the free energy of the hydrogen-adsorbed monolayers. Figure\,\ref{fig:site}(a)-(d) shows the free energy pathways of HER at pH$=0$. After full relaxation, hydrogen atoms are all adsorbed at the top sites, except for the chemically saturated atoms on the inter-fullerene bonds. All symmetry-irreducible adsorption sites of H$^{+}$ ions on qTP and qHP monolayers are listed in Figure\,\ref{fig:site}(e) and (f) respectively, numbered from low to high free energy. Due to the buckled structure of qHP, the sites on the convex (sites 2,\,5) and concave (sites 3,\,4) surfaces are inequivalent with different free energies of reaction intermediates. 

The hydrogen evolution reaction has two possible catalytic pathways\,\cite{Danilovic2012}: (1) Two protons are individually adsorbed onto the monolayer first, and then chemically desorb by combining into a hydrogen molecule (Volmer-Tafel, V-T), see Figure\,\ref{fig:site}(g); (2) A single proton is adsorbed first, and a second proton approaches the adsorbed proton to form an electrochemically desorbed hydrogen molecule (Volmer-Heyrovsky, V-H), see Figure\,\ref{fig:site}(h). The reaction scheme is summarised as
\begin{align*}
    \text{Volmer-Tafel:}\ &2(\mathrm{H}^{+}\mathrm{e}^{-}) + {}^* \rightarrow {}^*\mathrm{H}_2\\
    &{}^*\mathrm{H}_2 \rightarrow {}^* + \mathrm{H}_2\\
    \text{Volmer-Heyrovsky:}\ &(\mathrm{H}^{+}\mathrm{e}^{-}) + {}^* \rightarrow {}^*\mathrm{H}\\
    &(\mathrm{H}^{+}\mathrm{e}^{-}) +{}^*\mathrm{H} \rightarrow {}^* + \mathrm{H}_2
\end{align*}
where ${}^*$ is the catalyst, i.e. the C$_{24}$ monolayer, and $(\mathrm{H}^{+}\mathrm{e}^{-})$ denotes a pair of an aqueous proton in the solution and a photoexcited electron in the monolayer. 

The thermodynamic requirement for high reaction rates is that the free energy must decrease along each step in the reaction pathway. Among all the reaction intermediates of qTP through both the V-T and V-H mechanisms, even the lowest free energy barrier in Figure\,\ref{fig:site}(a) and (c) is still significantly larger than the thermal fluctuation energy $k_BT$ at room temperature (0.026\,eV) without photoexcitation. With photoexcitation, HER occurs spontaneously at all adsorption sites of qTP C$_{24}$ for $\text{pH}=0$ along both the V-T and V-H pathways, indicating promising photocatalytic efficiency. For qHP C$_{24}$, sites 1 and 2 are not active for the V-T mechanism because the free energies of their intermediates are lower than that of the final product, as shown in Figure\,\ref{fig:site}(b). However, HER at these two sites can still proceed simultaneously via the V-H route in Figure\,\ref{fig:site}(d). Furthermore, even up to neutral pH, the reactions at all available adsorption sites for both qTP and qHP monolayers are still spontaneous for at least one of the mechanisms. 

To investigate the microscopic mechanism of adsorption free energy, we compare the free energy at all sites with the bond angle strain, which is defined as the difference between the average of the three bond angles at each site and the standard $sp^2$ $120\degree$. The bond angle strain represents the deviation of the local carbon environment from an ideal flat $sp^2$ surface, i.e. graphene. As shown in Figure\,\ref{fig:site}(i), the adsorption free energy at all sites exhibits a strong correlation with the bond angle strain. This implies that the local bond environment has significant influence on hydrogen adsorption, especially for the small C$_{24}$ fullerene with relatively higher curvature compared to C$_{60}$.

Figure\,\ref{fig:site}(j) summarises the area density $\sigma_\mathrm{a}$ of thermodynamically active sites for qTP and qHP monolayers. Monolayer qHP C$_{24}$ has more surface active sites than qTP C$_{24}$ due to the presence of the additional site 1 on the single-bond sides of the C$_{24}$ units. At $\text{pH}=0$, C$_{24}$ monolayers have comparable active site densities to C$_{60}$ monolayers.  However, the number of active sites of C$_{60}$ monolayers decrease with increasing pH. Even at a moderate $\text{pH}=3$, the majority of the sites in C$_{60}$ monolayers are no longer active, while all the active sites in C$_{24}$ monolayers remain. Consequently, the number of surface active sites in C$_{24}$ monolayers is tripled compared to C$_{60}$ monolayers at $\text{pH}=3$, and the ratio continues to increase with higher pH. 

\section{Conclusion}

Using the smallest stable conventional [5,6]fullerene cage as building blocks, we predict two phases of monolayer C$_{24}$ networks with superior stability and strength comparing to their C$_{60}$ counterparts, indicating high feasibility in synthesising such monolayers with strong resilience to ambient temperatures and mechanical deformation.The band gaps of these C$_{24}$ monolayers are much larger than their C$_{60}$ counterparts while comparable to TiO$_2$, providing suitable band edges for photocatalytic water splitting in a wide pH range from 0 to 7. Additionally, both phases have strong optical absorption benefiting from multiple bright excitons from visible to UV-A range, enabling effective generation of a large amount of photoexcited carriers. Comparing to monolayer polymeric C$_{60}$, the density of surface active sites are tripled in C$_{24}$ monolayers. These results indicate that the photocatalytic performance of C$_{24}$ monolayers can be significantly enhanced.Beyond photocatalysis, we demonstrate the possibility of tuning physical and chemical properties of carbon nanomaterials by using different fullerene building blocks. Given the rich family of currently known fullerene molecules, new 2D materials can be designed with tunable and tailored functions. 

\section*{Acknowledgement}
J.W. acknowledges support from the Cambridge Undergraduate Research Opportunities Programme and from Peterhouse for a Bruckmann Fund grant and the James Porter Scholarship. D.Y. acknowledges support from Clare College for the Undergraduate Travel \& Research Grant. B.P. acknowledges support from Magdalene College Cambridge for a Nevile Research Fellowship. The calculations were performed using resources provided by the Cambridge Service for Data Driven Discovery (CSD3) operated by the University of Cambridge Research Computing Service (\url{www.csd3.cam.ac.uk}), provided by Dell EMC and Intel using Tier-2 funding from the Engineering and Physical Sciences Research Council (capital grant EP/T022159/1), and DiRAC funding from the Science and Technology Facilities Council (\url{http://www.dirac.ac.uk} ), as well as with computational support from the UK Materials and Molecular Modelling Hub, which is partially funded by EPSRC (EP/T022213/1, EP/W032260/1 and EP/P020194/1), for which access was obtained via the UKCP consortium and funded by EPSRC grant ref EP/P022561/1.

\end{document}